\documentclass{emulateapj}

\usepackage{mathrsfs}

\newcommand{\eg}{e.g., }
\newcommand{\ie}{i.e., }

\def\gsim{\mathrel{\rlap{\lower 4pt \hbox{\hskip 1pt $\sim$}}\raise 1pt
\hbox {$>$}}}
\def\lsim{\mathrel{\rlap{\lower 4pt \hbox{\hskip 1pt $\sim$}}\raise 1pt
\hbox {$<$}}}

\newcommand{\me}{m_{\rm e}}

\newcommand{\dpdp}[2]{\frac{\partial #1}{\partial #2}}
\newcommand{\dbdpdp}[2]{\frac{\partial^2 #1}{\partial #2^2}}

\newcommand{\bmm}[1]{\mbox{\boldmath $#1$}}

\slugcomment{Accepted for publication in the Astrophysical Journal Supplement Series.}

\begin{document}

\title{Time-dependent Multi-group Multidimensional Relativistic Radiative
Transfer Code Based On Spherical Harmonic Discrete Ordinate Method}

\author{
 Nozomu~Tominaga\altaffilmark{1,2},
 Sanshiro~Shibata\altaffilmark{3,1},
 Sergei~I.~Blinnikov\altaffilmark{4,5,2}
 }
\altaffiltext{1}{Department of Physics, Faculty of Science and
Engineering, Konan University, 8-9-1 Okamoto,
Kobe, Hyogo 658-8501, Japan; tominaga@konan-u.ac.jp}
\altaffiltext{2}{Kavli Institute for the Physics and Mathematics of the
Universe, The University of Tokyo, 5-1-5 Kashiwanoha, Kashiwa, Chiba
277-8583, Japan}
\altaffiltext{3}{Theory Center, Institute of Particle and Nuclear
 Studies, KEK, 1-1 Oho, Tsukuba 305-0801, Japan; sshibata@post.kek.jp}
\altaffiltext{4}{Institute for Theoretical and  Experimental Physics (ITEP),
Moscow 117218, Russia; Sergei.Blinnikov@itep.ru}
\altaffiltext{5}{VNIIA, 22 Sushchevskaya, 127055 Moscow, Russia}

\setcounter{footnote}{3}

\begin{abstract}
 We develop a time-dependent multi-group multidimensional relativistic radiative
 transfer code, which is required to numerically investigate
 radiation from relativistic fluids involved in, \eg gamma-ray bursts and active
 galactic nuclei. The code is based on the spherical harmonic discrete
 ordinate method (SHDOM) that evaluates a source function including
 anisotropic scattering in spherical harmonics and implicitly solves the
 static radiative transfer equation with a ray tracing in discrete
 ordinates. We implement treatments of time dependence, multi-frequency
 bins, Lorentz transformation, and elastic Thomson and inelastic Compton scattering to the
 publicly available SHDOM code. Our code adopts a mixed frame
 approach; the source function is evaluated in the comoving frame whereas
 the radiative transfer equation is solved in the laboratory frame. This
 implementation is validated with various test problems and
 comparisons with results of a relativistic Monte Carlo code. These
 validations confirm that the code correctly calculates intensity and its
 evolution in the computational domain. The code enables us to obtain an Eddington
 tensor that relates first and third moments of intensity (energy density
 and radiation pressure) and is frequently used as a closure relation in
 radiation hydrodynamics calculations. 
\end{abstract}


\section{INTRODUCTION}
\label{sec:intro}

Radiative transfer is an important piece of physics to describe how an
astronomical object is observed. Also the radiation often affects dynamical
behavior of the astronomical object. Thus it has been studied in many
astrophysical fields. However, the radiative transfer equation is intrinsically a
6-dimensional Boltzmann equation that is computationally expensive.
Therefore, the radiative transfer equation is frequently solved
in a simplified and approximated form appropriate for an object
of interest. 

Methods for the multidimensional radiative transfer and radiation
hydrodynamics are developed in various fields, \eg cosmological structure
formation \citep[\eg][for a review]{ili06,ili09}, star formation
\citep[\eg][]{kur07,tomida13}, a stellar and solar
atmosphere \citep[\eg][]{asp00,nor09}, and a terrestrial atmosphere
\citep[\eg][]{clo05,col06rt}. These methods are optimized for individual
research fields and involve various sophisticated physics for individual
phenomena, \eg a chemical network for cosmological structure formation
and star formation,
fine-structure lines for a stellar and solar atmosphere, and molecular
lines and scattering by dust for a terrestrial atmosphere. On the other hand,
they ignore terms with higher orders than $\mathcal{O}(v/c)$,
\ie they assume that a fluid velocity is much slower than the light velocity. This is a
commonly-used appropriate assumption to simplify the radiative transfer equation but
inapplicable for the radiative transfer in a relativistic flow.

An emission from a relativistic flow recently attracts researchers with
the advent of a gamma-ray burst (GRB).\footnote{The emission from a
relativistic flow is also 
interesting for, \eg active galactic nuclei (AGN).} The GRB is a
phenomenon emitting $\gamma$-ray photons from relativistic jets in a
short period and one of the brightest objects in the Universe. The
GRBs have been detected in the distant Universe with redshift as high as
$z\sim8.2$ (\citealt{tan09,sal09}; $z\sim9.4$ for a photometric
redshift, \citealt{cuc11}) and are believed to probe the high-$z$
Universe as well as quasars and galaxies. Furthermore, interestingly, it is
observationally exhibited that the $\gamma$-ray emission has
correlations of spectral peak energy versus isotropic radiation energy
\citep{ama02} and spectral peak energy versus peak luminosity
\citep{yon04}. A correlation between X-ray luminosity at the break
time and the break time is also suggested \citep{dai08}. These
correlations make researchers think of that the GRB can be a
standardizable candle being detectable at higher redshift than Type Ia
supernova \citep[\eg][]{ama08}.

The mechanism of GRB prompt emission and the origin of correlations have been
intensively studied. Observationally, many satellites and telescopes
report large variations of the prompt emission, for example, spectra with thermal
\citep[\eg][]{ryd10}, non-thermal \citep[\eg][]{abd09GRB080916C,zha09},
and high-energy components \citep[\eg][]{abd09GRB090902B,fan13},
polarization \citep[\eg][]{yon11,ueh12}, and duration \citep[\eg ultra-long
GRBs,][]{str13,lev14}. A correlation between optical and $\gamma$-ray
light curves is also exhibited \citep[\eg][]{ves05,woz09,gor12}. Theoretically, the
emission mechanism has been investigated by analytic studies
\citep[\eg][and references therein]{mes06} or numerical studies with
various assumptions: \eg superposing blackbody radiation from a
scattering photosphere \citep{bli99,miz11,nag11,laz13},\footnote{The
production site of photons is much deeper than the scattering
photosphere in a relativistic flow \citep[\eg][]{shibata14}.} solving a radiative
transfer equation in a spherical steady flow \citep{bel11}, transferring
photons in a steady flow with a relativistic Monte Carlo method
(\citealt{gia06,pee08,bel11,itogrb13,lun13}; S.~Shibata \& N.~Tominaga in prep.),
and calculating a spherical relativistic radiation hydrodynamics
\citep{tol05,tol13}. 

In spite of plenty observations and theoretical studies for many years, 
the mechanism of the GRB prompt emission is
still under debate \citep[\eg][]{zha14}. This is mainly because most of
studies is restricted to be qualitative and there are few quantitative
studies taking account of a structure of relativistic jets. In order to
investigate the GRB prompt emission quantitatively, a multidimensional
relativistic radiation hydrodynamics calculation is essentially
required. This is because the GRB is a relativistic and multidimensional
phenomenon and the radiation, of which energy can dominate the matter
energy, closely couples with matter. Therefore, it is
necessary to develop a radiative transfer code optimized for the GRBs
which fully includes the terms higher than $\mathcal{O}(v/c)$.

Much progress is recently made on the multidimensional radiation hydrodynamics
calculation, for example, (1) special relativistic 3-dimensional radiation
magnetohydrodynamics \citep{tak13,tak13b}, (2) general relativistic
3-dimensional radiation hydrodynamics \citep{sad13}, (3) 3-dimensional
radiation magnetohydrodynamics \citep{jia12,dav12,jia14}, (4) relativistic Monte Carlo
transport coupled with hydrodynamics \citep{rot15}, and (5) 3-dimensional
special relativistic Boltzmann hydrodynamics \citep{nag14}. However, the
calculations (1) and (2) are based on the M1 closure method, which can treat anisotropic
radiation field but not intersecting radiation from various
sources. This is not suitable for the GRBs because the material, that 
is surrounding the relativistic jet, \eg a cocoon, is hot and emits thermal
photons to varying directions. The calculation (3) adopts the variable Eddington tensor (VET) method
that can treat intersecting radiation from multiple sources and non-local
radiation equilibrium, but ignores terms with orders higher than
$\mathcal{O}(v/c)$. Recently, \cite{jia14} implements a radiative
transfer code involving time dependence and velocity dependent
source terms but the method is still accurate up to
$\mathcal{O}(v/c)$. The calculation (4) couples a relativistic Monte Carlo
transport method implicitly with hydrodynamics solvers and the
calculation (5) is, in particular, adopted for the neutrino transport in
a collapsing massive star. Although the calculations (4) and (5) would be
applicable for the GRBs, the calculation (4) can involve a noise of the
Monte Carlo even with a reducing technique they developed and the
calculation (5) is time-consuming and might be difficult to increase the
number of mesh points because it involves an inversion of a huge matrix. 
There are also general relativistic radiative transfer codes to describe
emission from surroundings of black holes
\citep[\eg][]{dex09,shc11}. The codes first derive photon geodesic
trajectories in curved spacetime and integrate a relativistic radiative
transfer equation along the geodesic paths. They also would be
applicable for the GRBs. However, they waste time and computational
resources because it is only required to calculate the
trajectory in the flat spacetime in the GRBs.

In this paper, we develop an implicit time-dependent multi-group
multidimensional special relativistic radiative transfer (RRT) code with
the mixed-frame approach. The RRT code can calculate an Eddington tensor
with the intensity and thus could be the first step to a time-dependent
special relativistic multidimensional multi-group radiation
hydrodynamics code optimized for the GRBs.
The RRT code is based on the spherical harmonic discrete
ordinate method (SHDOM) and takes into account ray tracing, time
dependence, Lorentz transformation, elastic Thomson and inelastic
Compton scattering. We present numerical test problems with the RRT code. 

We describe the method in Section~\ref{sec:method} and present test
problems in Section~\ref{sec:test}. The results are summarized in
Section~\ref{sec:summary}.

\section{Method}
\label{sec:method}

The RRT code is based on Spherical Harmonic Discrete Ordinate Method (SHDOM,
\citealt{eva98,pin09})\footnote{\url{http://nit.colorado.edu/shdom.html}}
code which is a publicly-available static monochromatic radiation
transfer code solving the following equation with the $\Lambda$
iteration (Picard iteration) method.
\begin{equation}
  \bmm{n}\cdot\nabla I_\nu(s,\bmm{n})= -\chi_\nu(s,\bmm{n})
  I_\nu(s,\bmm{n}) + \eta_\nu(s,\bmm{n}), \label{eq:shdom}
\end{equation}
where $s$ is a total path along a ray, \bmm{n} is direction of travel of
the photon, and $I_\nu$, $\chi_\nu$, and $\eta_\nu$ are an intensity, an
extinction coefficient, and an emission coefficient at frequency $\nu$,
respectively. Here, the extinction coefficient is the net absorption
coefficient of $\alpha_\nu+\sigma_\nu$, where $\alpha_\nu$ and
$\sigma_\nu$ are absorption and scattering coefficients, respectively. 
The SHDOM code is originally developed for radiative transfer
in the terrestrial atmosphere and thus correctly treats anisotropic
source terms including emission and scattering with spherical
harmonics. Equation~(\ref{eq:shdom}) is integrated for each ray 
described with discrete ordinates ($\theta$, $\phi$) with short characteristic
method (ray tracing). The SHDOM code transforms an intensity and a source function
between spherical harmonics and discrete ordinates at every
time step. The scheme of the SHDOM code, including the transformation
between spherical harmonics and discrete ordinates, the ray tracing, and
so on, is comprehensively described and validated in \cite{eva98} and \cite{pin09}. 

In order to apply the SHDOM code to a special relativistic phenomenon,
especially a GRB, the following processes and physics are necessary to
be included: (1) time dependence because a time step is needed to be 
short enough to capture a relativistic fluid that moves as fast as the
light speed, \ie the light-crossing time is as long as the
characteristic dynamical time, (2) Lorentz
transformation between laboratory and comoving frames which results in
relativistic beaming and a variation of photon frequencies, and (3)
anisotropic and inelastic Compton scattering because the photon energy is comparable with the
electron rest-mass energy, thus the assumption of Thomson scattering is
not valid, and the scattering dominates the opacity in the relativistic
jets of GRBs \citep[\eg][]{bel13}. 

We set a computational domain to be translationally symmetric along the
$y$ axis but the ray of radiation is solved in three dimension and
described with polar coordinates $\theta$ and $\phi$. The coordinates
are set to have zenith direction of the $z$ axis and azimuth angle
measured from the $x$ axis to the $y$ axis.
The radiative transfer equation is solved with the mixed-frame approach
\citep[\eg][]{mih84,hub07}; the photon ray is traced in the laboratory frame and 
the source term is evaluated in the comoving
frame. In addition to the above
implementation, we update the ray tracing scheme to a cubic Bezier
interpolant method \citep{del13} and include an acceleration scheme of
the $\Lambda$ iteration (Ng acceleration, \citealt{ng74}), but omit the
adaptive treatment of mesh points and the parallelization
with Message Passing Interface (MPI) for simplicity in this paper. 

\subsection{Time dependence}
\label{sec:time}

A time-dependent radiative transfer equation is written as
\begin{eqnarray}
\nonumber {1\over{c}}\dpdp{I_\nu(t,s,\bmm{n})}{t}&+&\bmm{n}\cdot\nabla
 I_\nu(t,s,\bmm{n}) \\
 &=& -\chi_\nu(t,s,\bmm{n}) I_\nu(t,s,\bmm{n}) + \eta_\nu(t,s,\bmm{n}),
\end{eqnarray}
where $c$ is the light speed.
A finite difference approximation to the time derivative is written as 
\begin{equation}
 \dpdp{I_\nu(t,s,\bmm{n})}{t}={I_\nu(t+\Delta t,s,\bmm{n})-I_\nu(t,s,\bmm{n})\over{\Delta t}}.
\end{equation}
The time-dependent radiative transfer equation is deformed to the same
shape as the static radiative transfer equation (Eq.~\ref{eq:shdom})
with modified absorption and emission coefficients as
\begin{equation}
 \bmm{n}\cdot\nabla I_\nu(t+\Delta t,s,\bmm{n}) = -\tilde{\chi}_\nu I_\nu(t+\Delta t,s,\bmm{n}) +
 \tilde{\eta}_\nu, \label{eq:time}
\end{equation}
where 
\begin{eqnarray}
 \tilde{\chi}_\nu &=& \chi_\nu(t+\Delta t,s,\bmm{n}) + {1\over{c\Delta t}}\\
 \tilde{\eta}_\nu &=& \eta_\nu(t+\Delta t,s,\bmm{n}) + {I_\nu(t,s,\bmm{n})\over{c\Delta t}}.
\end{eqnarray}
Since $I_\nu(t,s,\bmm{n})$ and $\Delta t$ are given, the equation
(\ref{eq:time}) can be implicitly solved with the original SHDOM code. 
Such an implementation of time dependence has been successfully
adopted in previous studies \citep{bar09,hil12,jac12}. 
We note that the 4th order Runge-Kutta scheme is adopted to proceed the
time step (Appendix~\ref{appendix}). 

\subsection{Lorentz transformation}

Variables at the next time step are evaluated in the laboratory frame with
discrete ordinates with the ray tracing method. The variables are first
Lorentz transformed with discrete ordinates to the comoving frame before the conversion from
the discrete ordinates to the spherical harmonics. Then, the source terms are
evaluated in the comoving frame with spherical harmonics. The source
terms are first converted from spherical harmonics to discrete
ordinates and then Lorentz transformed from the comoving frame to the
laboratory frame. The obtained source term is adopted for the ray tracing in
the laboratory frame.

The Lorentz transformation from the laboratory frame to the comoving
frame moving with $\bmm{v}$, corresponding to the Lorentz factor of
$\Gamma$, is defined with the following equations \citep[\eg][]{mih84}
\begin{eqnarray}
 \nu_0&=&\Gamma\nu\left(1-{\bmm{n}\cdot\bmm{v}\over{c}}\right)
 \label{eq:lorentzfreq} \\
 \bmm{n}_0&=&{\nu\over{\nu_0}}\left\{\bmm{n}-\Gamma{\bmm{v}\over{c}}\left[1-{\Gamma\over{\Gamma+1}}{\bmm{n}\cdot\bmm{v}\over{c}}\right]\right\}, \label{eq:lorentzray}
\end{eqnarray}
where physical values in the comoving frame are denoted with a subscript
$0$. 

The Lorentz transformation is required for an intensity $I$, an emission
coefficient $\eta$, and an extinction coefficient $\chi$, whose
Lorentz invariants are $I/\nu^3$, $\eta/\nu^2$, and $\chi\nu$,
respectively. The frequency and traveling direction of radiation are 
transformed with Equations (\ref{eq:lorentzfreq}) and
(\ref{eq:lorentzray}).\footnote{The subroutine is taken from
\url{http://cernlib.web.cern.ch/cernlib/version.html}.}
We prepare $(\theta,\phi)$ mesh points in the laboratory and comoving frames
and the Lorentz invariants along the rays, which are transformed from a
frame and remapped to the rays in the other frame, and then the
intensity, emission coefficient, and absorption coefficient are obtained
in the other frame.

\subsection{Thomson and Compton scattering}

\begin{figure*}
\epsscale{1.}
\plotone{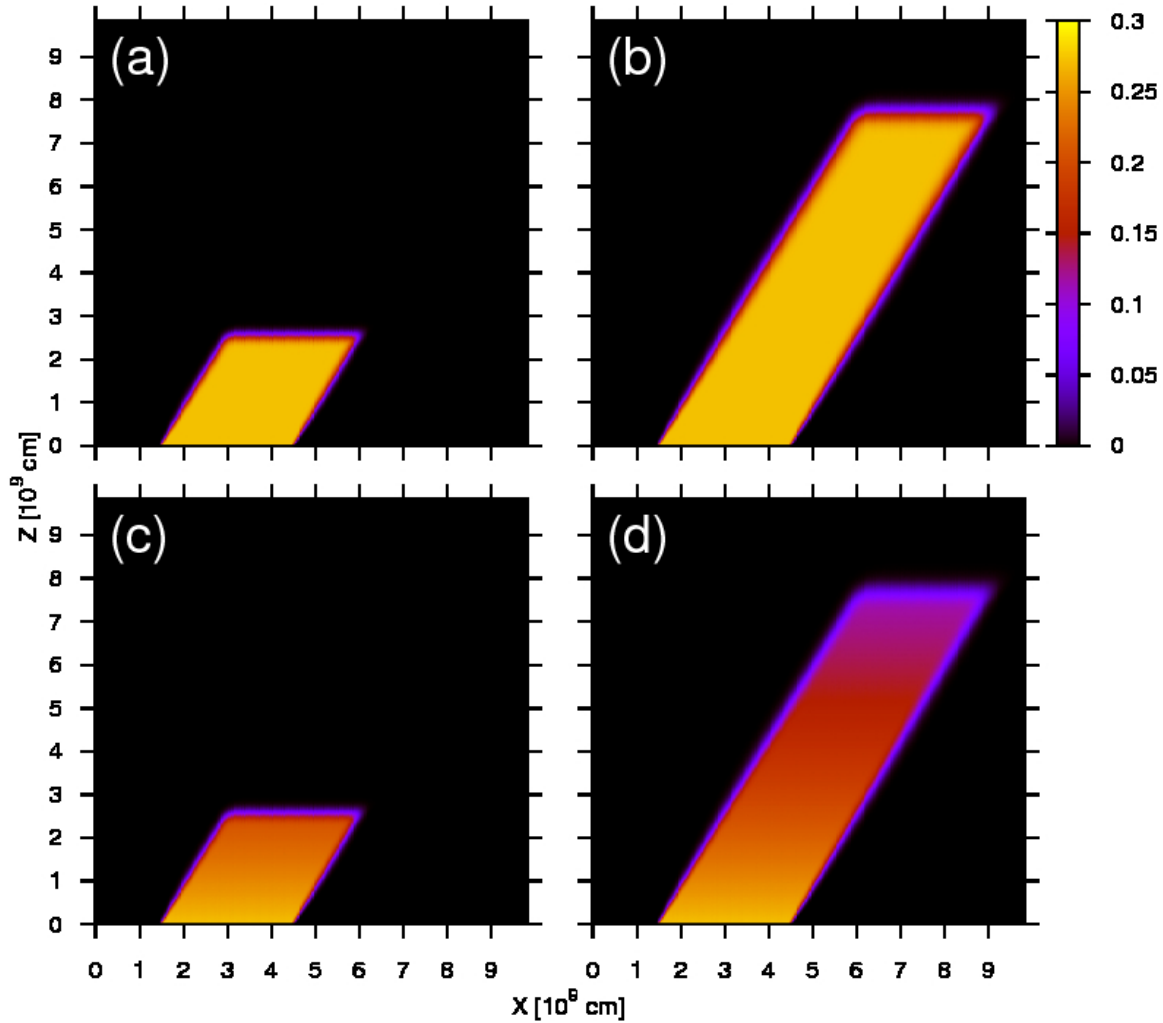}
\figcaption{Snapshots of mean intensity $J$ for
 (a) $t=0.1$~s and $\alpha=0$, (b) $t=0.3$~s and $\alpha=0$, (c)
 $t=0.1$~s and $\alpha=10^{-10}$~cm$^{-1}$, and (d) $t=0.3$~s and
 $\alpha=10^{-10}$~cm$^{-1}$. The medium is at rest ($v=0$) and the $x$
 and $z$ axes are divided by $512$ mesh points. \label{fig:timeevo2}}
\end{figure*}

The source function due to scattering is evaluated with spherical
harmonics as the original SHDOM code does. Since the prescription is
expatiated in \cite{eva93,eva98}, we briefly describe the procedure.
The source function $S(\mu,\phi)$ is expanded in spherical harmonics
space as
\begin{equation}
 S(\mu,\phi)=\sum_{lm} Y_{lm}(\mu,\phi) S_{lm},
\end{equation}
where $Y_{lm}(\mu,\phi)$ are orthonormal real-valued spherical harmonics
functions, whereas the phase function of the scattering ${d\sigma\over{d\Omega}}$ is
expanded in a Legendre series in the scattering angle as
\begin{equation}
 {d\sigma\over{d\Omega}} = \sum^{N_L}_{l=0} \chi_l \mathcal{P}_l,
\end{equation}
where $N_L$ and $\mathcal{P}_l$ are the maximum order of Legendre polynomials,
and Legendre polynomials, respectively. The source function is computed as
\begin{equation}
 S_{lm} = {\omega \chi_l\over{2l+1}}I_{lm} + T_{lm},
\end{equation}
where $\omega$ is the single scattering albedo, and $I_{lm}$ and
$T_{lm}$ are intensity and thermal emission expanded in spherical
harmonics, respectively.

We adopt the phase functions of Thomson and Compton scattering.
The phase function of Thomson scattering is 
\begin{equation}
 {d\sigma_{\rm T}\over{d\Omega}}={r_0^2\over{2}}\left(1+\cos^2\Theta\right),
\end{equation}
where $r_0$ is the classical electron radius and $\Theta$ is the
scattering angle \citep{ryb85}.
Klein-Nishina scattering differential cross section is adopted for the
Compton scattering. The equations are 
\begin{eqnarray}
 \nu_1&=&{\nu\over{1+{h\nu\over{\me c^2}}\left(1-\cos\Theta\right)}}\\
 {d\sigma\over{d\Omega}}&=&{r_0^2\over{2}}{\nu_1\over{\nu}}\left({\nu\over{\nu_1}}+{\nu_1\over{\nu}}-\sin^2\Theta\right),
 \label{eq:comptoncross}
\end{eqnarray}
where $\nu_1$ is the photon frequency after the scattering and 
$\me $ is the rest mass of electron \citep{ryb85}.

A change of the photon frequency and dependence of the scattering kernel
on the photon frequency are essential features of the Compton
scattering. Therefore, we implement a multi-group treatment for test
calculations of the Compton scattering. The differential cross section
${d\sigma\over{d\Omega}}|^{(i),(j)}$ of an incident photon in $i$-th
frequency bin with a range of $\left[\nu^{(i)},\nu^{(i)}+\Delta\nu^{(i)}\right]$ to
$j$-th frequency bin with a range of 
$\left[\nu_1^{(j)},\nu_1^{(j)}+\Delta\nu_1^{(j)}\right]$ is a
frequency-dependent scattering kernel defined by Equation
(\ref{eq:comptoncross}). ${d\sigma\over{d\Omega}}|^{(i),(j)}$ is
expanded in a Legendre series in the scattering angle with frequency-dependent
single scattering albedo. In the mixed-frame approach, the photon exchange between
frequency bins takes place only in the comoving frame. 

\begin{figure}
\epsscale{1.}
\plotone{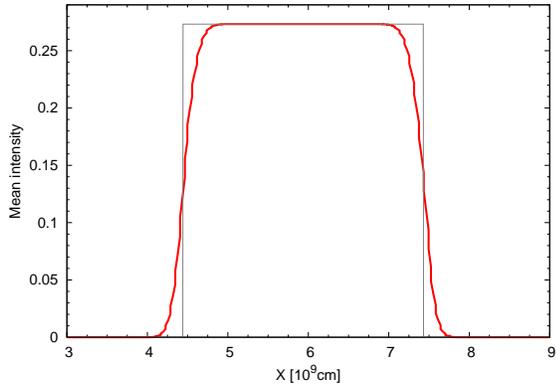}
\figcaption{Profile section at $z=5.0\times10^{9}$~cm for the case of
 $\alpha=0$ at $t=0.25$~s (red) and an analytical result (gray).
 \label{fig:timecut}}
\end{figure}

\section{Test Problems}
\label{sec:test}

\cite{eva98} and \cite{pin09} have intensively tested the original
SHDOM code and investigated its efficiency, accuracy, and scalability. Therefore, we focus on
validation tests of time dependence, Lorentz transformation, and
Thomson and Compton scattering in this paper.

\subsection{Searchlight beam test}

A searchlight beam test has been performed in various studies (\eg
\citealt{ric01,tur01,tak13,tak13b}). A narrow beam of light is introduced into
the computational domain at a certain angle and the beam crosses the domain. This
test examines whether a code can solve a time-dependent radiative 
transfer equation and how radiation disperses along the path.

We set a computational domain of $10^{10}$cm square,
of which $x$ and $z$ axes are divided by $512$ mesh points. Numbers of
angular mesh points are
$(N_{\theta},N_{\phi})=(4,8)$. The origin is the
bottom left corner of the domain. A beam of light is injected from the
bottom boundary at 
$1.5\times10^{9}~{\rm cm}<x<4.5\times10^{9}~{\rm cm}$ along a ray with
an angle of $(\theta,\phi)=(0.17\pi,0)$. Two
cases of absorption coefficients are tested; the domain is uniformly
filled with a medium with $\alpha=0$ or $\alpha=10^{-10}$~cm$^{-1}$.

Figures~\ref{fig:timeevo2}(a)-\ref{fig:timeevo2}(d) show snapshots of the
mean intensity $J$ in the domain at $t=0.1$~s and $0.3$~s. The beam crosses 
the domain properly with time at the injected angle. Figure~\ref{fig:timecut} shows
the profile section at $z=5.0\times10^{9}$~cm and $t=0.25$~s for the
case of $\alpha=0$. The
radiation disperses slightly laterally because the RRT code adopts the short
characteristic method.

\begin{figure}
\epsscale{1.}
\plotone{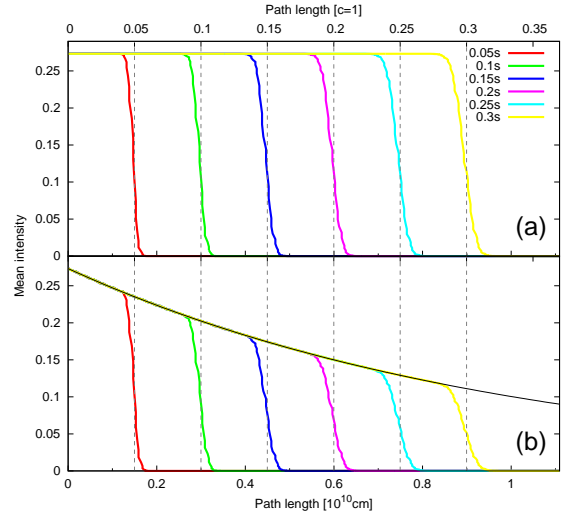}
\figcaption{Time evolution of the mean intensity $J$ along a ray with
 $x=\tan\left(0.17\pi\right)z+3.0\times10^{9}~{\rm cm}$ at $t=0.05$~s (red), $0.1$~s
 (green), $0.15$~s (blue), $0.2$~s (magenta), $0.25$~s (cyan), and
 $0.3$~s (yellow) for the cases of (a) $\alpha=0$ and (b)
 $\alpha=10^{-10}$~cm$^{-1}$. The vertical dashed lines represent analytical
 expectations of the wave front proceeding with $c$ at $t=0.05$, $0.1$,
 $0.15$, $0.2$, $0.25$, and $0.3$~s (gray dashed). The
 analytical result of radiation attenuation is also shown (black).
 \label{fig:timeevo}}
\end{figure}

Figures~\ref{fig:timeevo}(a)-\ref{fig:timeevo}(b) show $J$
along $x=\tan\left(0.17\pi\right)z+3.0\times10^{9}~{\rm cm}$. 
The figures demonstrate that a wave front proceeds with the light
speed. However, the wave front is smeared. This is because the
time dependence is taken into account with $\tilde{\chi}$ and
$\tilde{\eta}$ which exponentially reduce and/or enhance the intensity as a
function of a path length $s$. The widths of the smooth profile at the wave
front are identical for the cases of $\alpha=0$ and
$\alpha=10^{-10}$~cm$^{-1}$. The mean intensity of the beam is constant
for the case of $\alpha=0$ and is reduced in accordance with
$\exp\left(-\alpha s\right)$ for the case
of $\alpha=10^{-10}$~cm$^{-1}$. The test confirms that the time
dependence and the radiation attenuation are correctly solved by the
RRT code although the wave front is smeared.

\begin{figure*}
\epsscale{1.}
\plotone{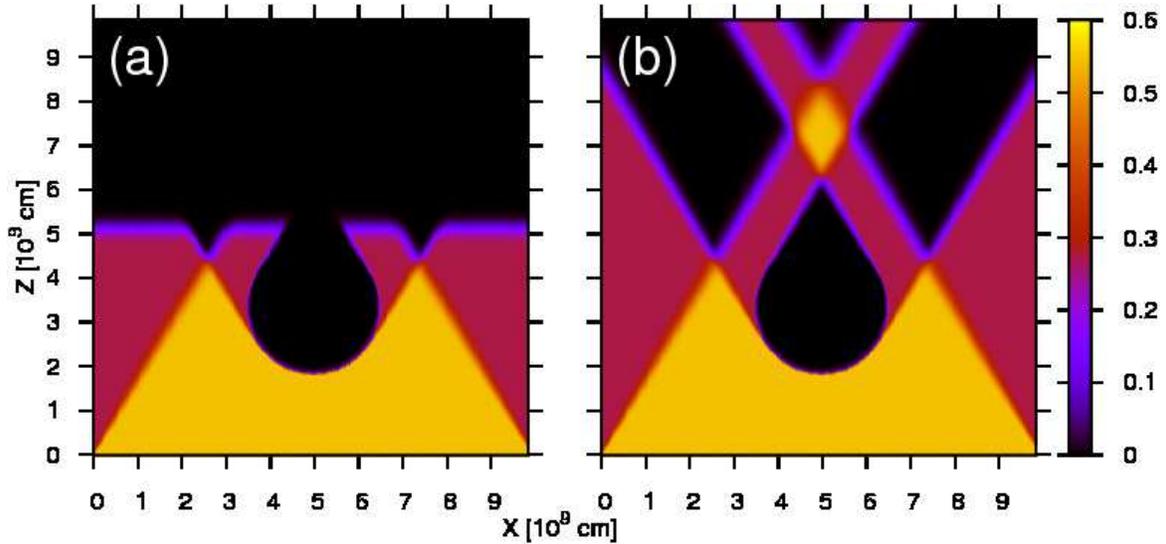}
\figcaption{Snapshots of the mean intensity $J$ for the two beam and
 shadow test at (a) $t=0.2$~s and (b)
 $1.0$~s. The light is injected from the bottom boundary at the 
 angles with $(\theta,\phi)=(0.17\pi,0)$ and $(0.17\pi,\pi)$. An
 optically thick cylinder with
 $\alpha=3.3\times10^{-8}~{\rm cm}^{-1}$ and a radius of
 $1.5\times10^9$~cm is located at 
 $(x,z)=(5.0\times10^9~{\rm cm}, 3.3\times10^9{\rm cm})$. The medium is
 at rest ($v=0$) and the $x$ and $z$ axes are divided by $512$
 mesh points. \label{fig:twobeamshadow}}
\end{figure*}

\subsection{Two beam shadow test}

A shadow test was proposed in \cite{hay03}. This test examines a
reproduction of shadow behind an optically thick blob when 
plane-parallel radiation illuminates the blob. It is required to
properly take into account at least up to the first moment of intensity
in order to reproduce the shadow. Thus, for instance, a
method with a diffusion approximation fails this test. 

On the other hand, a two beam test has been performed in \eg
\cite{dav12,jia12,sad13,jia14}. This test examines whether two independent
beams proceeding at different angles pass through without any
interactions when they intersect. To describe this
phenomenon, it is required to properly take into account at least up to
the second moment of intensity. Thus, approximate schemes
with a closure relation treating up to the first moment, \eg the M1 closure
method, fail this test. 

We solve a two beam with shadow test combining the above tests, which is
performed in \cite{dav12,jia12,sad13,jia14}. We set an optically thin
($\alpha=0$) computational domain of $10^{10}$cm square, where $x$ and
$z$ axes are divided by $512$ mesh points. Number of angular mesh points are
$(N_{\theta},N_{\phi})=(4,8)$. The origin is the
bottom left corner of the domain. An optically thick absorptive cylinder
with a radius of $1.5\times10^9$cm is located at
$(x,z)=(5.0\times10^9~{\rm cm},3.3\times10^9~{\rm cm})$ perpendicular to
the $xz$ plane.
The optical depth of the cylinder is set to be $\tau=100$ with the
diameter, \ie $\alpha=3.3\times10^{-8}$~cm$^{-1}$.
Plane-parallel radiation is injected from the bottom boundary at angles
of $(\theta,\phi)=(0.17\pi,0)$ and $(0.17\pi,\pi)$.

Figures~\ref{fig:twobeamshadow}a-\ref{fig:twobeamshadow}b show snapshots
of $J$ in the domain at $t=0.2$ and $1.0$~s. The
plane-parallel radiation proceeds properly with time and the wave
speed is the speed of light. A shadow develops behind the cylinder along the
directions of two beams and the two beams cross around 
$(x,z)=(5.0\times10^9~{\rm cm},7.4\times10^9~{\rm cm})$ without any
interaction. The test confirms that the calculation properly solves the
radiation transfer equation for intensity.

\begin{figure*}
\epsscale{1.}
\plotone{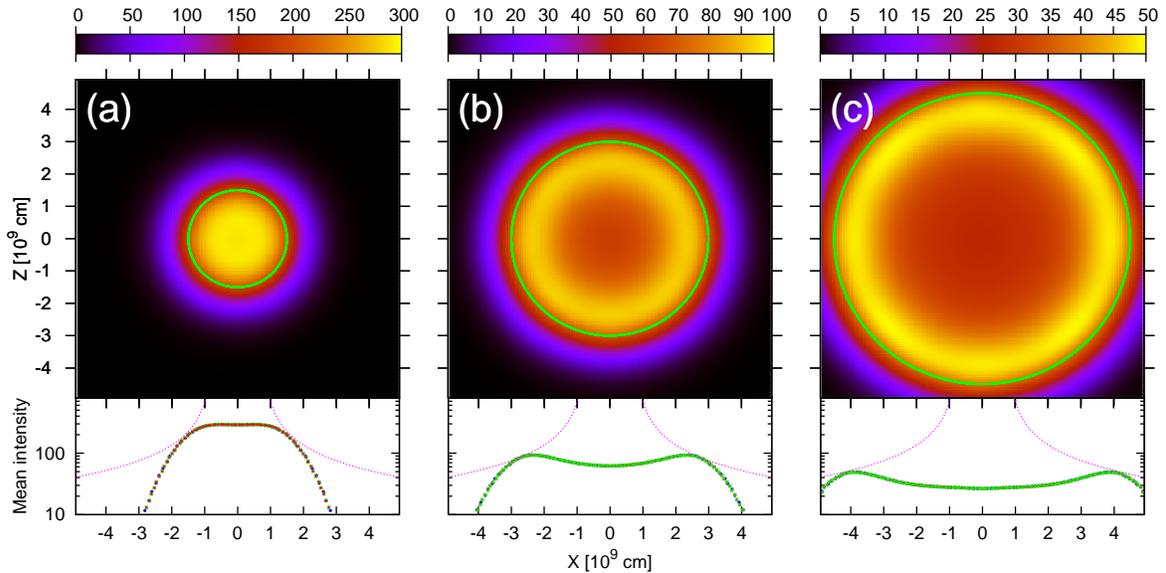}
\figcaption{(Top) Snapshots of the mean intensity $J$ for the optically thin
 ($\chi=0$)
 radiative pulse test at $t=0.1$~s (left), $0.2$~s (middle), and $0.3$~s
 (right). (Bottom) Profile section of the mean intensity along the $x$ axis
 (red), $z$ axis (green), and $x=z$ (blue) at $t=0.1$~s (left), $0.2$~s
 (middle), and $0.3$~s (right). The $1/r$ dependence is also shown in the
 bottom panels (magenta). \label{fig:pulse}}
\end{figure*}

\begin{figure}
\epsscale{1.}
\plotone{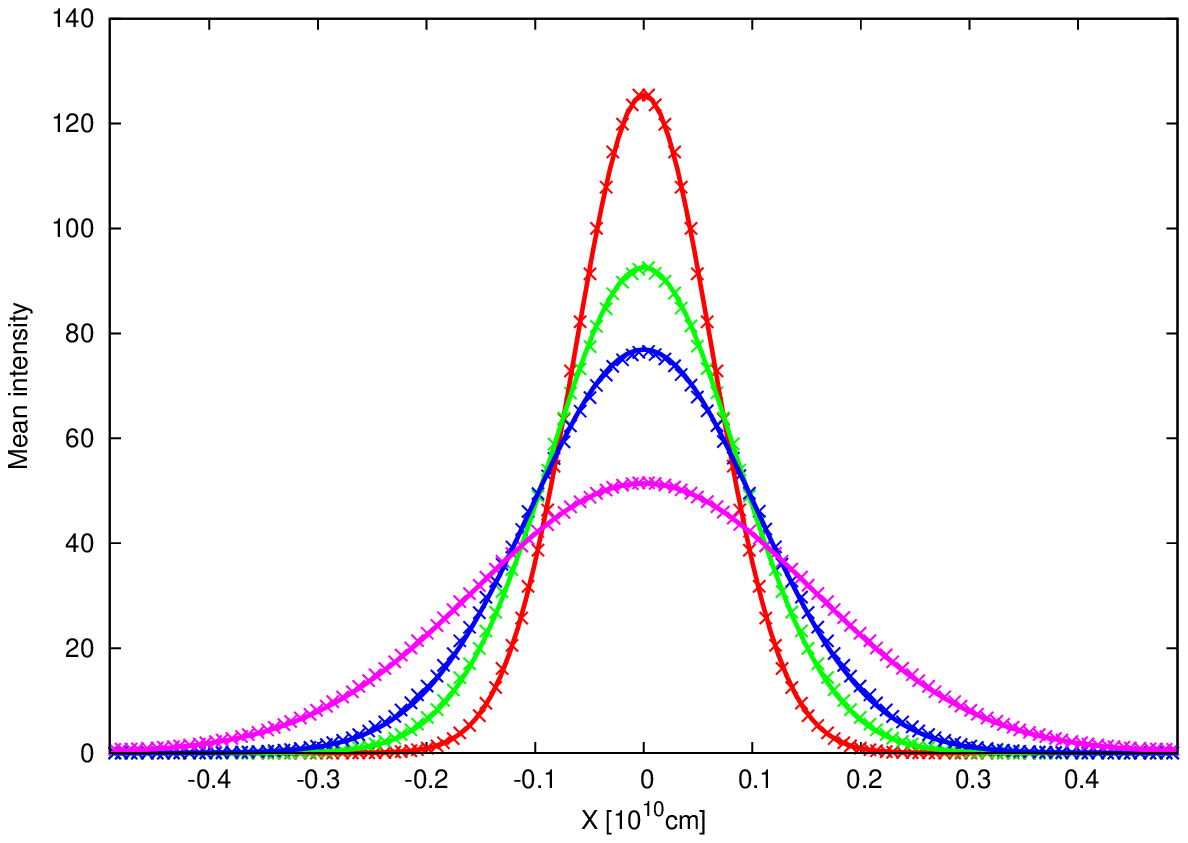}
\figcaption{Profile section of the mean intensity $J$ along the $x$ axis
 (crosses) for the optically thick ($\sigma=3\times10^{-8}~{\rm cm^{-1}}$)
 radiative pulse test at $t=0$ (red), $0.5$~s (green),
 $1.0$~s (blue), and $3.0$~s (magenta). The solid lines show the exact
 solution (Eq.~\ref{eq:exactdiffusion}). \label{fig:diffusion}}
\end{figure}

\subsection{Radiative pulse test}

The RRT code is tested with the evolution of a radiative pulse initially
having a Gaussian profile in an optically thin medium and a
scattering-dominated optically thick medium. The tests have been
performed in \cite{sad13} and verify the treatments of time dependence
and scattering.

\subsubsection{Optically thin medium}

In the optically thin limit, the isotropic radiative pulse spreads
with the speed of light and the mean intensity decreases inversely
proportionally to the radius in the translational symmetry.

We set an optically thin computational domain of $10^{10}$cm square
with $\alpha=0$. The origin is at the center of the domain. Each axis is
divided by $128$ mesh points. Number of angular
mesh points are $(N_{\theta},N_{\phi})=(64,128)$. An initial radiative pulse
is set according to the equation
\begin{equation}
 I_0(\bmm{x},\bmm{n})=100 \exp\left[-\left({r\over{w}}\right)^2\right],
\end{equation}
where $r~(=|\bmm{x}|)$ is the distance from the origin and
$w=9.0\times10^8$cm. The intensity of the pulse is initially isotropic
at each mesh point.

The top panels of Figure~\ref{fig:pulse} show
snapshots of $J$ at $t=0.1$~s, $0.2$~s, and $0.3$~s. The
expected size of the pulse is also shown by a green circle with a radius
of $ct$. 
The bottom panels of Figure~\ref{fig:pulse} show $J$ along $x$ axis, $z$
axis, and a line with $x=z$ at
$t=0.1$~s, $0.2$~s, and $0.3$~s. These profiles of $J$ are identical and
the peak of $J$ decreases with
the expansion of the pulse according to $J\propto 1/r$ as expected. 
The result demonstrates that the pulse isotropically propagates with
the speed of light and that the geometrical dilution of radiation
is correctly followed with the RRT code.

\begin{figure*}
\epsscale{1.}
\plotone{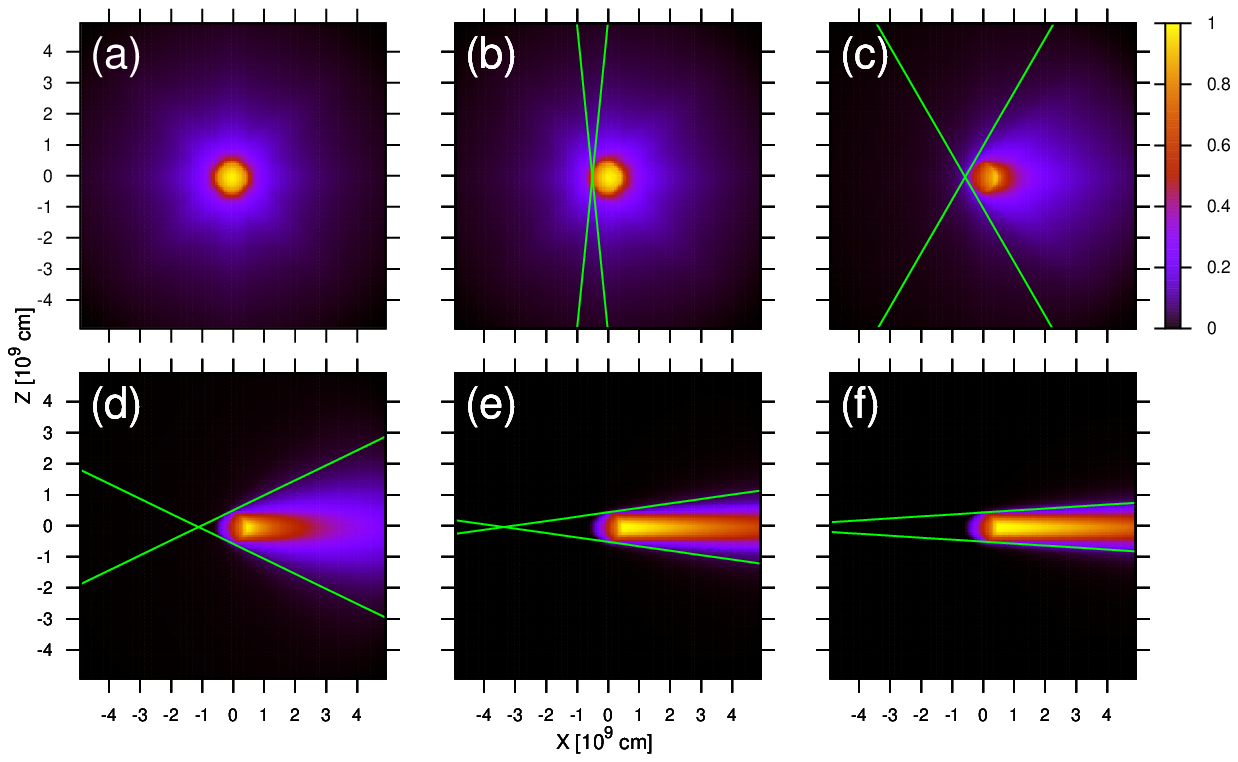}
\figcaption{Snapshots of the mean intensity normalized with the maximum
 mean intensity $J/J_{\rm max}$ for the relativistic beaming
 test for the cases of (a) $(v_x,v_z)=(0,0)$, (b) $(v_x,v_z)=(0.1c,0)$
 (c) $(v_x,v_z)=(0.5c,0)$, (d) $(v_x,v_z)=(0.9c,0)$, (e) $(v_x,v_z)=(0.99c,0)$, and
 (f) $(v_x,v_z)=(0.995c,0)$. Asymptotic analytic expressions of the beaming effect
 are also shown (green lines).\label{fig:lorentz}}
\end{figure*}

\subsubsection{Optically thick medium}

In the scattering-dominated optically thick medium, the radiative pulse
diffuses out. A one-dimensional diffusion equation
$\dpdp{J}{t}=\chi\dbdpdp{J}{x}$ can be solved analytically and the
solution is written as
\begin{equation}
 J(t,x)={1\over{2\sqrt{\chi\pi t}}}\int_{-\infty}^{\infty} J_0(x')
 \exp\left\{-{\left(x-x'\right)^2\over{4\chi t}}\right\} dx'
 \label{eq:exactdiffusion}
\end{equation}
where $\chi={1\over{3c\sigma}}$ and $J_0(x)$ is the initial profile of the
radiative pulse.

We set the optically thick computational domain of $10^{10}$cm square
with $\tau=300$ along a side, \ie $\sigma=3\times10^{-8}~{\rm cm^{-1}}$.
The origin is set at the center of the domain. Each axis is divided by 
$128$ mesh points. Number of angular mesh points are
$(N_{\theta},N_{\phi})=(32,64)$. The each mesh point is optically thick,
$\tau=2.3$, so that most of the photons are scattered at least once in
the mesh point. Here, we adopt a spherical scattering kernel. A radiative
pulse is initially set according to  
\begin{equation}
 I_0(\bmm{x},\bmm{n})=100 \exp\left[-\left({x\over{w}}\right)^2\right]
\end{equation}
where $w=9.0\times10^8$cm. The intensity of the pulse is initially
isotropic at each mesh point.

Figure~\ref{fig:diffusion} shows the time evolution of $J$
along the $x$ axis. The radiation gradually diffuses in the numerical
simulation and the distribution of $J$ well reproduces
that of the analytic solution at every time step until the radiation
reaches the boundaries of the computational domain. 

\begin{figure}
\epsscale{1.}
\plotone{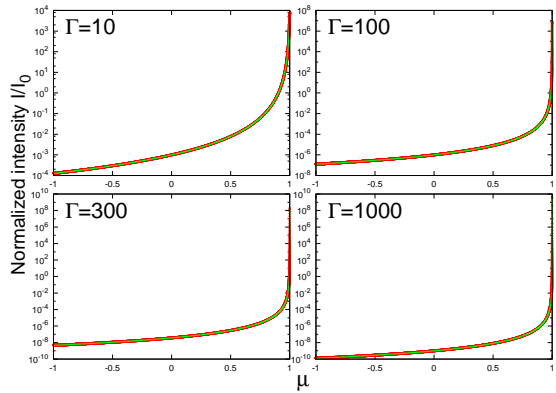}
\figcaption{Distribution of the normalized intensity $I/I_0$ as a function of
 the zenith angle for the numerical solution (red) and the analytic
 solution (green). \label{fig:lorentzpart}}
\end{figure}

\subsection{Relativistic beaming test}

We validate the ability of the RRT code to treat the relativistic beaming
as a result of the Lorentz transformation. This is a characteristic
feature of a special relativistic radiative transfer
calculation. Although the propagation of beam of light along a curved
trajectory is demonstrated in general relativistic radiative transfer
calculations \citep[\eg][]{sad13}, this is the first attempt to confirm
the relativistic beaming in special relativistic calculations.

We set a computational domain of $10^{10}$~cm square filled with the
optically thin medium with $\alpha=0$ and consider a cylindrical light
source with a radius of $R=5\times10^8$cm at the center of a
computational domain. The center of a
computational domain is set to be the origin.
In the cylinder, particles move with $\bmm{v}=(v_x,v_z)$ with respect to
the laboratory frame and emit photons isotropically in the comoving
frame of each particle, whereas they do not emit any photons and not
interact with photons outside the cylinder. Each axis is divided by
$128$ mesh points and number of angular mesh points are
$(N_{\theta},N_{\phi})=(32,64)$. 

Figures~\ref{fig:lorentz}a-\ref{fig:lorentz}f show snapshots of the
normalized mean intensity $J/J_{\rm max}$
in the Laboratory frame at $t=0.2$~s for the cases with 
$(v_x,v_z)=(0,0)$, $(0.1c,0)$, $(0.5c,0)$, $(0.9c,0)$, $(0.99c,0)$, and
$(0.995c,0)$, respectively, where $J_{\rm max}$ is the
maximum mean intensity in the computational domain. The corresponding Lorentz factors are 
$\Gamma=1$, $1.005$, $1.15$, $2.29$, $7.09$, and $10.0$. It is clearly shown
that the mean intensity is high along the direction of $\bmm{v}$. An
analytic expression of the beaming effect (asymptotically $\theta\sim{1\over{\Gamma}}$,
\citealt{ryb85}) is also shown with green lines circumscribing the light
source. The biased distribution of $J$ in the calculation is
consistent with the analytic expression. The test shows that the beaming effect is
correctly taken into account in the RRT code.

The speed of $0.995c$ is close to the maximum speed that can be
correctly solved with $N_{\theta}=32$ because the half-angle of the
concentration of radiation for the case of $v=0.9952c$ is comparable to
the interval between angular mesh points. Although the test is limited
due to computational resources, the Lorentz factor as
high as the one encountered in GRB models can be resolved by the RRT code, if the larger number of
angular mesh points is adopted, because the mapping subroutine for the
Lorentz transformation is valid even for high Lorentz factor.
Figure~\ref{fig:lorentzpart} shows the angular distribution of the
intensity $I$ in the laboratory frame for the cases with
$(v_x,c-v_z)=(0,5\times10^{-3}c)$ ($\Gamma=10$), $(0,5\times10^{-5}c)$ ($\Gamma=100$),
$(0,6\times10^{-6}c)$ ($\Gamma=300$), and $(0,5\times10^{-7}c)$ ($\Gamma=1000$),
which is transformed from an isotropic intensity $I_0$ in the comoving
frame. The deviation from the analytical solution
$I(\mu)=\left({{\Gamma (1-\mu v_z/c)}}\right)^{-3}I_0$ is less than
$10^{-6}I$.  Here, we adopt the large number of angular mesh points of
$(N_{\theta},N_{\phi})=(4096,8192)$. We note that the mapping subroutine correctly
transforms an isotropic radiation for any Lorentz factor even with the
small number of angular mesh points.

\begin{figure*}
\epsscale{1.}
\plotone{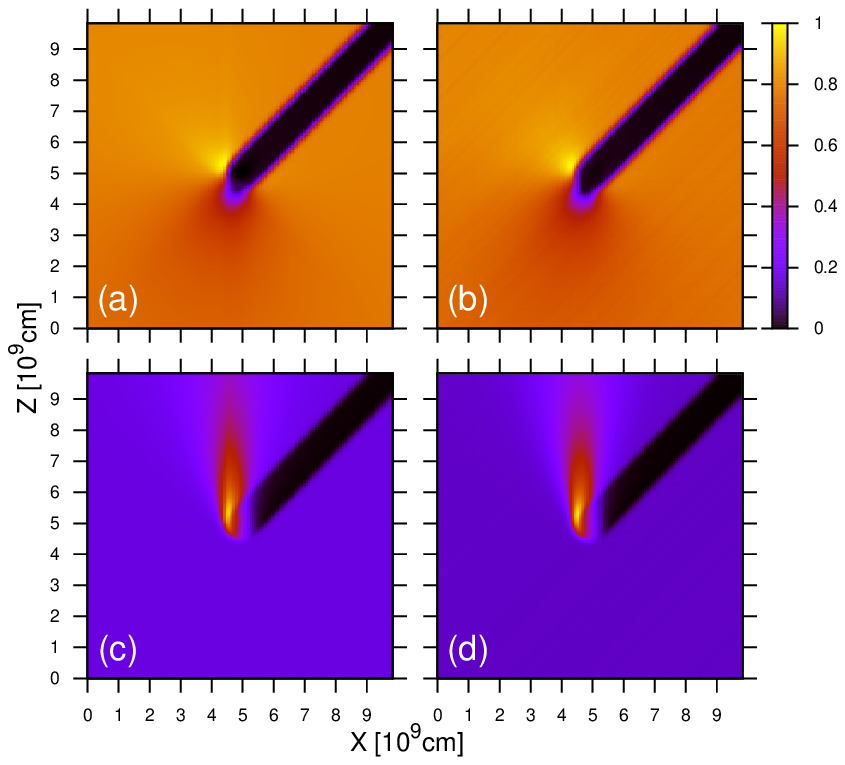}
\figcaption{Snapshots of $z$-oriented fluxes normalized with the maximum
 flux $F_z(x,z)/F_{z,{\rm max}}$ in the medium with
  $(v_x,v_z)=(0,0)$ [(a) and (b)] and $(v_x,v_z)=(0,0.9c)$ [(c) and
 (d)]. The panels show the results of the RRT code [(a) and (c)] and the
 RMC code [(b) and (d)]. Here, the Thomson
 scattering kernel is adopted and the plane-parallel radiation is injected
 from the left and bottom boundaries at a single angle of
 $(\theta,\phi)=(0.24\pi,0)$.
 \label{fig:thomson}}
\end{figure*}

\subsection{Comparison with Monte Carlo method}

The original SHDOM code and the Monte Carlo method have been carefully
compared and their drawback and advantage have been presented in
\cite{eva98} and \cite{pin09}. 
Therefore, we test the RRT code only on the treatments of the Thomson and Compton
scattering and Lorentz transformation, and the implementation of
multiple frequency groups for the Compton scattering tests, by
comparing solutions of a shadow test of the RRT code with those of a
relativistic Monte Carlo (RMC) code (S.~Shibata \& N.~Tominaga in
prep., Appendix B for a test of the RMC code). In this subsection, we omit the time
dependence and adopt the static version of the RRT code because the
RMC code currently does not follow the time evolution of the
radiation. 

Here, we adopt an ideal test problem. A scattering-dominated optically
thick cylinder with $\sigma=\sigma_0 \exp\left(-{r^2\over{w^2}}\right)$,
where $\sigma_0=2\times10^{-7}~{\rm cm^{-1}}$ and $w=2.5\times10^8$~cm,
is put at the center of the optically thin computational domain of
$10^{10}$~cm square with $\sigma=2\times10^{-15}~{\rm cm^{-1}}$. 
Scattering particles
rest or flow with $\bmm{v}$ in the scattering cylinder.\footnote{We
assume, for example, an ideal atom that do not interact with the photons
if they are neutral and that the atoms rest or flow with $\bmm{v}$ in
the computational domain and its ionization fraction changes to
reproduce the distribution of $\sigma$.} 
Plane-parallel radiation is injected from the left and bottom boundaries
at a single angle of $(\theta,\phi)=(0.24\pi,0)$. 
The incident angle of radiation is set to the same in the RRT and 
RMC calculations.

\begin{figure*}
\epsscale{.8}
\plotone{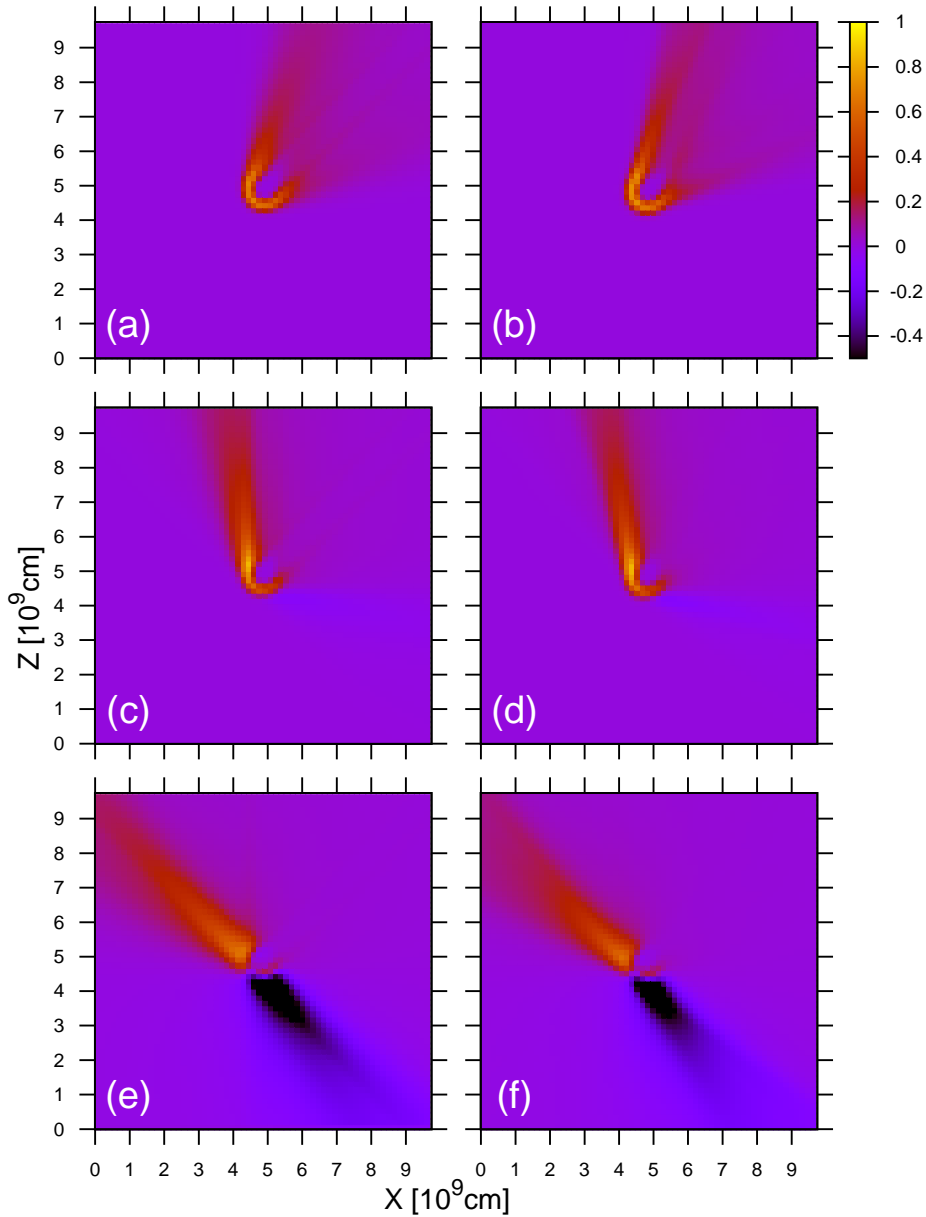}
\figcaption{Snapshots of frequency-dependent $z$-oriented fluxes normalized with the maximum
 flux $F_{\nu,z}(x,z)/F_{\nu,z,{\rm max}}$ in the cases with
  $(v_x,v_z)=(0,0)$. The panels represent fluxes at the energy bins with
 $\left[0.88\me c^2,0.99\me c^2\right]$ [(a) and (b)],
 $\left[0.66\me c^2,0.77\me c^2\right]$ [(c) and (d)],
 and $\left[0.44\me c^2,0.55\me c^2\right]$ [(e) and (f)]. The panels
 show the results of the RRT code [(a), (c), and (e)] and the RMC code
 [(b), (d), and (f)]. Here, the frequency-dependent
 Compton scattering kernels are adopted.  \label{fig:compton_v0}}
\end{figure*}

\begin{figure*}
\epsscale{.8}
\plotone{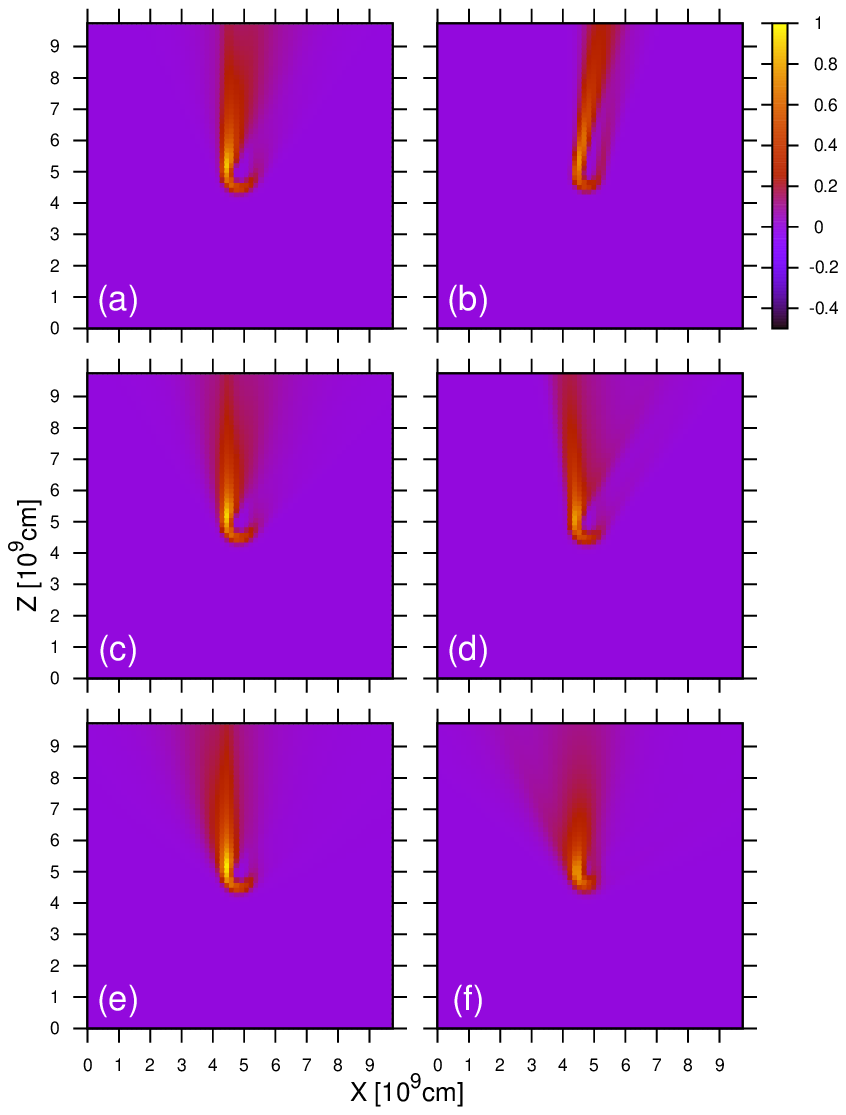}
\figcaption{Same as Figure~\ref{fig:compton_v0} but for the cases with
  $(v_x,v_z)=(0,0.9c)$. The panels represent fluxes at the energy bins with
 $\left[2.0\me c^2,2.4\me c^2\right]$ [(a) and (b)],
 $\left[1.2\me c^2,1.6\me c^2\right]$ [(c) and (d)],
 and $\left[0.4\me c^2,0.8\me c^2\right]$ [(e) and (f)]. \label{fig:compton_vz3}}
\end{figure*}

\subsubsection{Thomson scattering}

In this subsection, we adopt a kernel of Thomson scattering.
The computational domain is divided by $128\times128$ mesh points and numbers of
angular mesh points are $(N_{\theta},N_{\phi})=(160,320)$. 

Figures~\ref{fig:thomson}(a)-\ref{fig:thomson}(b), and
\ref{fig:thomson}(c)-\ref{fig:thomson}(d) show normalized $z$-oriented
fluxes $F_z(x,z)/F_{z,{\rm max}}$ obtained by the RRT and RMC calculations in the cases
with $(v_x,v_z)=(0,0)$ and $(0,0.9c)$,
respectively, where $F_{z,{\rm max}}$ is the
maximum $z$-oriented flux in the computational domain. Figures~\ref{fig:thomson}(a)-\ref{fig:thomson}(b)
demonstrate that the $z$-oriented fluxes are small below the cylinder
due to the scattered photons propagating to the $-z$ direction, whereas the
$z$-oriented fluxes are high at the top left side of the cylinder. Since
the number density of scattered photons decreases with the distance from
the scattering cylinder as $1/r$, the $z$-oriented fluxes are
prominently modified around the cylinder. Also
the shadow behind the cylinder consistently forms in both
calculations. Figures~\ref{fig:thomson}(c)-\ref{fig:thomson}(d) shows that 
the scattered photons are concentrated to the $+z$ direction due to the
relativistic beaming in both calculations. These demonstrate
that the RRT code well solves the anisotropic scattering and the Lorentz
transformation.

\subsubsection{Compton scattering}

In this subsection, we adopt frequency-dependent kernels of Compton
scattering and the multi-group treatment.
The computational domain is divided by $64\times64$ mesh points. We
test the RRT code for 2 cases with different velocity of scattering
particles with $(v_x,v_z)=(0,0)$ and $(0,0.9c)$.
Numbers of angular mesh points are $(N_{\theta},N_{\phi})=(128,256)$. The
frequency range is equally divided by $10$ bins in
both cases but the maximum frequency is $h\nu=1.1\me c^2$ for the
cases with $(v_x,v_z)=(0,0)$ and $4.0\me c^2$
for the case with $(v_x,v_z)=(0,0.9c)$. Monochromatic light with
$h\nu=1.05\me c^2$ for the cases with 
$(v_x,v_z)=(0,0)$ and $h\nu=1.0\me c^2$ for the
case with $(v_x,v_z)=(0,0.9c)$ is injected. 

Figures~\ref{fig:compton_v0}(a)-\ref{fig:compton_v0}(f) and 
\ref{fig:compton_vz3}(a)-\ref{fig:compton_vz3}(f) 
show frequency-dependent normalized $z$-oriented fluxes $F_{\nu,z}(x,z)/F_{\nu,z,{\rm max}}$ in the cases with $(v_x,v_z)=(0,0)$ and 
$(0,0.9c)$, respectively, where $F_{\nu,z,{\rm max}}$ is the
maximum $z$-oriented flux in a frequency bin in the computational domain. Each panel shows the fluxes in different energy
bins. There was no light in these energy bins before
scattering. This is the reason why no shadow appears behind the
cylinder. The negative fluxes due to the scattered photons produce the
black region below the cylinder in Figure~\ref{fig:compton_v0}(e) and
\ref{fig:compton_v0}(f). Although the shape of $F_{\nu,z}(x,z)$ is smeared in
the RRT code, especially in the case of $(v_x,v_z)=(0,0.9c)$, because the
frequency of photons is converted to the central frequency of each bin
at every time step, it provides the similar snapshots of the
$z$-oriented flux in the RMC code. These
display that the RRT code correctly solves the Compton scattering and the Lorentz
transformation, and well treats the multi-group radiation transfer
equation.

\section{Summary}
\label{sec:summary}

We develop a time-dependent multi-group multidimensional relativistic
radiative transfer code by implementing the treatments of time
dependence, multi-frequency bins, Lorentz transformation, and elastic
Thomson and inelastic Compton scattering to the publicly available SHDOM
code. The SHDOM code evaluates a source
function in spherical harmonics and solves a static radiative transfer
equation with a ray tracing in discrete ordinates. The RRT code is
validated by the various tests and the comparison with the RMC
calculations. The searchlight beam, two beam with shadow,
radiative pulse, and relativistic beaming tests are successfully passed
by the RRT code and confirm that the RRT code correctly handles the time
dependence and the Lorentz transformation. The results of the RRT code are
consistent with those of the RMC code and the comparisons
verify the implementation of elastic Thomson and inelastic Compton scattering
and multi-group treatment in the RRT code. The RMC code, in turn, is
validated against the EGS tools (Appendix~\ref{appendix2}).

The RRT code enables us to obtain the evolution of intensity and thus to
self-consistently derive an Eddington tensors without approximations
like the flux limited diffusion or the M1 closure methods. We emphasize
that the radiation tends to be more anisotropic in the relativistic
fluid because of the Lorentz transformation and Compton scattering for
the $\gamma$-ray photons, and
thus that the angular distribution of radiation should be properly taken
into account. Combining the Eddington tensors with relativistic
hydrodynamics calculations \citep[\eg][]{tom09a}, a relativistic
radiation hydrodynamics will be realized with the variable Eddington
tensor method. Furthermore, the RRT code
implicitly solves the radiative transfer equation and thus can follow
the radiative transfer in a non-relativistic fluid like a supernova
without adopting unnecessarily short time steps. Such a method will be
useful to clarify the connection between GRBs and supernovae. 

It is currently difficult to increase the numbers of frequency bins, and
angular and spatial mesh points due to an available memory
resource. The
difficulties are solved if the adaptive treatment of mesh points and the
parallelization with distributed memory are implemented
because the ray tracing in the laboratory frame and the evaluation of
source function in the comoving frame are independent of each ray and
each mesh point, respectively. However, we note the low resolution of the
frequency is an intrinsic drawback of the multi-group treatment compared to
the Monte Carlo method, in which a frequency of each photon changes
continuously. Thus, the Monte Carlo method is superior to the RRT code 
for spectral synthesis calculations of lines and fine spectral features.
However, the method intrinsically involves a noise and it is
expensive to reduce the noise because the reduction is realized only
proportional to the square root of the number of photon
packets.\footnote{Several techniques are suggested to reduce the noise
\citep[\eg][]{ste13,rot15}.} And, the large number of photon packets is
necessary to follow the time dependence because the photon packets are 
emitted at each time step. Furthermore, the radiation contributes to the
hydrodynamics with an integration of radiation over all frequency and
the dynamical effects of the radiation in GRBs are not dominated by
narrow spectral lines. Thus, we propose a post-processing Monte Carlo
calculation for spectral synthesis after a time-dependent matter-coupled
relativistic radiative transfer calculation.

\acknowledgements

N.T. thanks Hajime Susa for fruitful discussion.
This research has been supported in part by the
RFBR-JSPS bilateral program, World Premier
International Research Center Initiative, MEXT, Japan, and by the
Grants-in-Aid for Scientific Research of the JSPS (23740157, 15H05440).
S.B. is supported by the Russian Science Foundation Grant
No. 14-12-00203.
Calculations were in part carried out on the general-purpose PC farm
at Center for Computational Astrophysics, National Astronomical
Observatory of Japan. 

\appendix

\section{Time integration}
\label{appendix}

We treat the time dependence with the modified absorption and emission
coefficients as shown in Section~\ref{sec:time}, and implicitly obtain
the self-consistent time-dependent intensity at $t+\Delta t$. Although 
the time derivative is differentiated in the first order, we adopt the
4th order Runge-Kutta scheme to proceed the time step from $t$ to
$t+\Delta t$ by dividing the time interval to 4 steps, in order to
increase the accuracy by adopting smaller time intervals.

\begin{eqnarray}
I_\nu(t+\Delta t,s,\bmm{n}) = I_\nu(t,s,\bmm{n}) + 
 \Delta I_\nu^{(1)} + \Delta I_\nu^{(2)}
  + \Delta I_\nu^{(3)} + \Delta
  I_\nu^{(4)}, \label{eq:rk4th}
\end{eqnarray}
where
\begin{eqnarray}
 \Delta I_\nu^{(1)} &=& \mathscr{I}_\nu(\Delta t/6,I_\nu(t,s,\bmm{n}))-I_\nu(t,s,\bmm{n}) \\
 I_\nu^{(1)} &=& I_\nu(t,s,\bmm{n})+3\Delta I_\nu^{(1)}\\
 \Delta I_\nu^{(2)} &=& \mathscr{I}_\nu(\Delta t/3,I_\nu^{(1)})-I_\nu^{(1)} \\
 I_\nu^{(2)} &=& I_\nu(t,s,\bmm{n})+{3\over{2}}\Delta I_\nu^{(2)}\\
 \Delta I_\nu^{(3)} &=& \mathscr{I}_\nu(\Delta t/3,I_\nu^{(2)})-I_\nu^{(2)} \\
 I_\nu^{(3)} &=& I_\nu(t,s,\bmm{n})+3\Delta I_\nu^{(3)}\\
 \Delta I_\nu^{(4)} &=& \mathscr{I}_\nu(\Delta t/6,I_\nu^{(3)})-I_\nu^{(3)}.
\end{eqnarray}
Here, $\mathscr{I}_\nu(\Delta t,I_\nu)$ is the solution of
Equation (\ref{eq:time}) with $I_\nu$ and $\Delta t$, \ie
$\mathscr{I}_\nu(\Delta t,I_\nu(t,s,\bmm{n}))=I_\nu(t+\Delta t,s,\bmm{n})$.

\section{Comparison with electron gamma shower (EGS) software}
\label{appendix2}

\begin{figure}
\epsscale{.83}
\plotone{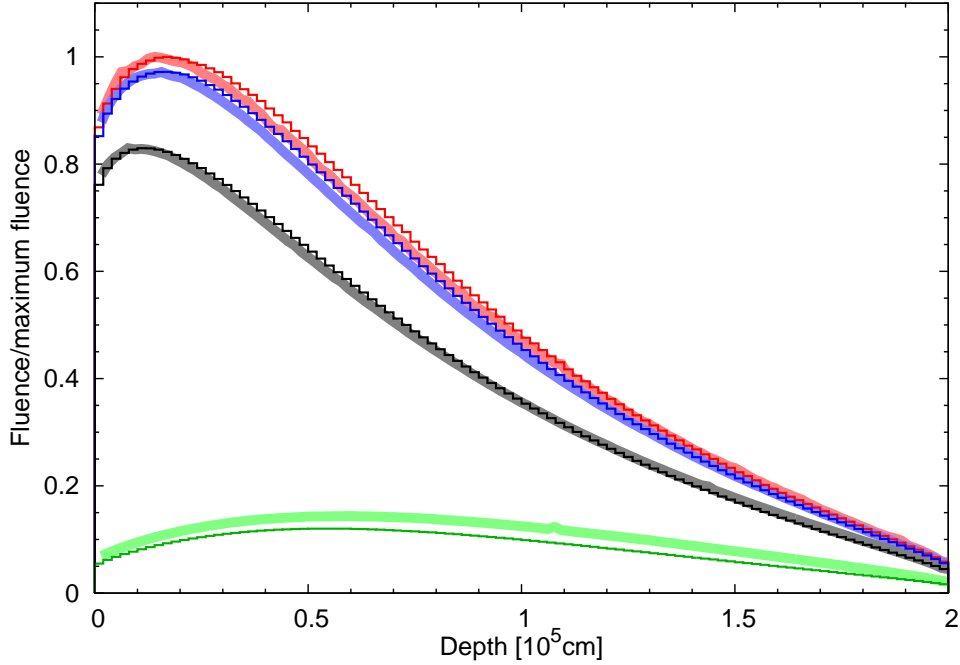}
\figcaption{Fluence as a function of the depth calculated with the
 RMC code (thick) and the EGS tool (thin). The
 color represents the rings with $r<1\times10^4$~cm (red), 
 $1\times10^4~{\rm cm}<r<5\times10^4$~cm (blue),
 $5\times10^4~{\rm cm}<r<1\times10^5$~cm (black), 
 and $1\times10^5~{\rm cm}<r<2\times10^5$~cm (green). \label{fig:egs}}
\end{figure}

We compare the result of the RMC code with that of National
Research Council's electron gamma shower (EGS) software
tool\footnote{\url{http://www.nrc-cnrc.gc.ca/eng/solutions/advisory/egsnrc\_index.html}} to
confirm its validity, especially the
treatment of the scattering.
EGS software is a publicly available code that enables sophisticated treatment of
photon, electron, and positron transfer in complicated medium and
provides graphic tools to show numerical results. Here, we adopt 
{\sl flurznrc} package only with the Compton scattering by electrons at
rest in the EGS software. 

We set a cylindrical computational domain with a radius of
$2\times10^5$~cm and a depth of $2\times10^5$~cm. The cylinder is filled
with H atoms with a number density of $8.37\times10^{-5}~{\rm g~cm^{-3}}$.
The incident photons with $50$~keV are vertically injected from the
top boundary with $r<1\times10^5$~cm, where $r$ is the distance from the
center of the top boundary. 

Figure~\ref{fig:egs} shows a comparison with the fluence as a function
of the depth in rings with $r<1\times10^4$~cm, 
$1\times10^4~{\rm cm}<r<5\times10^4$~cm, $5\times10^4~{\rm cm}<r<1\times10^5$~cm, 
and $1\times10^5~{\rm cm}<r<2\times10^5$~cm, normalized by the maximum
fluence. The fluence is
defined as a summation of $1/|\cos\theta|$
for all photons per unit area at each boundary of slabs, where $\theta$ is the angle
the photon makes with respect to the normal to the plane. We adopt
$6\times10^7$ photons for both of the RMC calculation and the
calculation by the EGS tool. The RMC codes give
consistent results with that obtained by the EGS software.

\end{document}